\documentclass[%
 reprint,
superscriptaddress,
 amsmath,amssymb,
 aps,
 pra,
floatfix,
]{revtex4-2}

\usepackage[normalem]{ulem} 
\usepackage{graphicx}
\usepackage{dcolumn}
\usepackage{bm}
\usepackage[dvipsnames]{xcolor}

\begin{document}

\preprint{APS/123-QED}

\title{Optical snake states in photonic graphene}

\author{O.M.~Bahrova}
\affiliation{Center for Theoretical Physics of Complex Systems, Institute for Basic Science (IBS), Daejeon 34126, Republic of Korea}
\author{S.V.~Koniakhin}
\affiliation{Center for Theoretical Physics of Complex Systems, Institute for Basic Science (IBS), Daejeon 34126, Republic of Korea}
\affiliation{Basic Science Program, Korea University of Science and Technology (UST), Daejeon 34113, Korea}
\author{A.V.~Nalitov}
\affiliation{Moscow Institute of Physics and Technology, Dolgoprudnyi 141701, Russia}
\affiliation{Faculty of Science and Engineering, University of Wolverhampton, Wulfruna Street, Wolverhampton WV1 1LY, United Kingdom}
\author{E.D.~Cherotchenko}
\affiliation{Ioffe Institute, St. Petersburg 194021, Russia}


\begin{abstract}
We propose an optical analogue of electron snake states based on artificial gauge magnetic field in  photonic graphene with effective strain implemented by varying distance between pillars.
We develop an intuitive and exhaustive continuous model based on tight-binding approximation and compare it with numerical simulations of a realistic photonic structure.
The allowed lateral propagation direction is shown to be strongly coupled to the valley degree of freedom and the proposed photonic structure may be used a valley filter.
\end{abstract}

\


\maketitle


Snake states were first theoretically proposed in 2D electron gases under spatially non-uniform magnetic fields as snake-like electron trajectories, emerging along the lines of the out-of-plane field component sign-switching~\cite{Muller1992, reijniers2000snake, Reijniers2002}.
Such peculiar unidirectionally guided states breaking the time-reversal symmetry found most promising applications in graphene, where confinement of massless electrons in potential traps is forbidden due to Klein paradox \cite{Oroszlny2008, Liu2015a,Ghosh2008,Zuo2022}.
Significance of snake states in graphene is further highlighted by their phenomenological similarity to topological edge states of quantum Hall effect, achievable even at room temperatures \cite{Novoselov2007}.
Finally, externally applied non-uniform magnetic field in graphene can be replaced with an effective field due to properly arranged mechanical strain \cite{Pereira2009}.

Certain graphene properties were reproduced in its photonic analogue, a honeycomb lattice etched out of a planar optical microcavity, including Dirac cones \cite{Jacqmin2014,Amo2019PRX, Lu2021} and Klein tunneling \cite{ Jiang2020, Konyakhin_Li_2022}.
Other features, such as effective pseudospin-orbit coupling \cite{ZChen2015, Nalitov2015,Shuang2019, Whittaker2021} and topological bands and edge states \cite{Rechtsman2013, Nalitov2015a,Klembt2018,Tang2023, Ren2023}, are specific to photonic graphene.
Aside from effective magnetic fields coupled to photon spin and stemming from planar cavity mode energy splitting, strain-induced Abelian synthetic gauge field coupled to momentum is also present in photonic graphene \cite{Jamadi2020,Mann2020, Huang2022}.

One of the most distinctive features of topologically nontrivial photonic graphene band structures is the emergence of protected from backscattering unidirectionally guided optical states.
Although this property, stemming from band topological nontriviliaty, relies on external time-reversal symmetry breaking, similar behaviour can be realized for photons with selected by quasi-resonant excitation~\cite{Kang2018}.
While in the former case the unidirectional transport can be employed in an optical insulator for information processing~\cite{Solnyshkov2018}, the latter feature can be used for valley splitting or valley filtering~\cite{Fujita2010}.
\begin{figure}
    \centering
    \includegraphics[width=7.5cm]{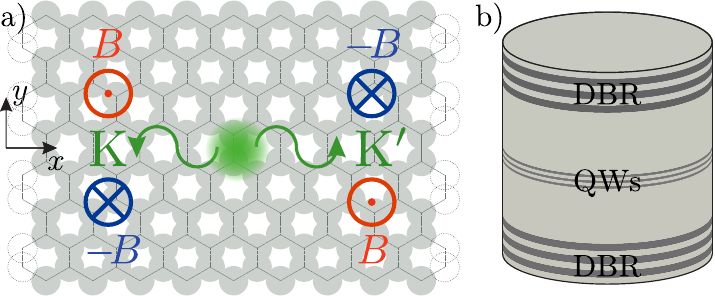}
    \caption{A sketch of the optical cavity structure and the proposed effect. a) Top view of the strained graphene lattice. Uniform along the $x$ axis,  structure is symmetrically deformed in the $y$ direction. Only vertical nearest-neighbor coupling is affected: it is locally equal to the uniform coupling in the zig-zag chain direction at $y=0$ and is gradually increasing toward the edges. This generates a normal pseudomagnetic field switching its direction between the two semiplanes and the two valleys K and K', producing optical snake states of opposite directions in the two valleys. b) Side view of a cavity pillar at each lattice vertex.  Cavity mode is confined between two parallel Distributed Bragg Reflectors and couples to excitons in the Quantum Wells.}
    \label{fig:fig0}
\end{figure}

In this letter we propose an optical analogue of electron snake states due to strain-induced spatially nonuniform synthetic magnetic fields emerging in photonic graphene ribbons with varying distance between adjacent cavity pillars.
We compare results of the semi-analytic continuous low-energy approximation model with that of numerical simulations based on both the tight-binding model and two-dimensional Schr\"odinger equation.
Finally, we discuss the valley filtering property of proposed snake states due to coupling of the effective synthetic magnetic field to the valley index and their relation to photonic topological edge states similar to the ones discussed in~\cite{Kang2018}.

Nonlinear mean-field dynamics of scalar photonic field is described with the Gross-Pitaevskii equation \cite{CarusottoRev2013}
\begin{equation}  \label{eq:GPE}
    i \hbar \partial_t \Psi = \left[ - \frac{\hbar^2}{2m} (\partial_x^2+\partial_y^2) + V(x,y) +  g |\Psi|^2 - \imath\frac{\gamma}{2}  \right] \Psi,
\end{equation}
with $m$ being the effective mass of the planar cavity mode.
In the linear ballistic propagation regime, neglecting the interaction and decay terms, governed by $g$ and $\gamma$ parameters, one recovers the 2D Schr\"odinger equation
\begin{equation} \label{eq:SE}
    i \hbar \partial_t \Psi = \left[ - \frac{\hbar^2}{2m} (\partial_x^2+\partial_y^2) + V(x,y) \right] \Psi.
\end{equation}

Here the shape of potential $V(x,y)$ is defined by the etching pattern in optical microcavity.
Observation of snake states requires contact of two semiplanes with opposite directions of magnetic (or pseudomagnetic as in our case) field.
Thus, we employ the strained optical graphene structure similar to the one studied in Ref.~\cite{Jamadi2020}, where photonic Landau levels were observed, supplemented with its reflected copy, see Fig.~\ref{fig:fig0}.
In the following, we first approach the model analytically, showing the similarity between electron snake states and their photonic counterparts, and then proceed by numerically simulating an experimentally relevant setup with a given effective potential profile $V(x,y)$.

Considering the nearest-neighbor approximation the Hamiltonian can be easily reduced to the one of photonic graphene in the vicinity of either of the two Dirac points (low-energy approximation) \cite{Nalitov2015}\footnote{See Supplementary Materials}:
\begin{equation} \label{eq:H0}
    H = \hbar v_F (\tau_z \sigma_x \hat{q}_x + \sigma_y \hat{q}_y ),
\end{equation}
where $\tau_z = \pm 1$ is the valley index specifying one of the two Dirac points ($K$ and $K'$), $\sigma_x$ and $\sigma_y$ are the Pauli matrices in the basis of
the two sublattices, and $\hbar v_F = 3aJ/2$ is the characteristic velocity with $a$ and $J$ being the distance and the coupling parameter between adjacent cavities of the lattice.
Spatial variation of the coupling parameter $J(y) = J(1 - |y|/\delta)$ results in an additional term $J|y|/\delta$, which is symmetric with respect to the axis $y=0$, in the Hamiltonian Eq.~\eqref{eq:H0}, and is equivalent to an effective vector potential $\mathbf{A}$:
\begin{equation}
    A_x = -\frac{2\tau_z|y|}{3a\delta}, \; A_y = 0,
\end{equation}
corresponding to an effective magnetic field $\mathbf{B} = \nabla \times \mathbf{A}$ normal to the lattice plane.
The effective field magnitude $B=2/(3a\delta)$ is uniform, while its direction is inverted at the symmetry axis separating the two half-spaces $y>0$ and $y<0$.

\begin{figure}[ht!]
    \centering
    \includegraphics[width=8.5cm]{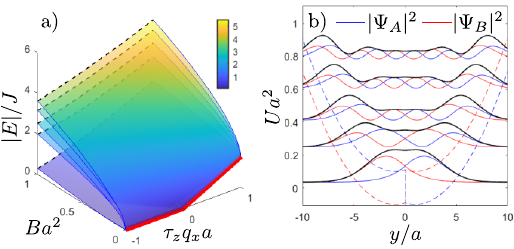}
    \caption{Continuous snake state model spectrum. a) Energy dispersion as a function of effective magnetic field $B$. The thick red line shows the classical limiting case $B=+0$. b) Effective Schr\"odinger equation spectra for both sublattices. Black lines show the total wave function density inheriting the system symmetry. Parameters: $Ba^2=-\tau_zq_xa=0.1$. }
    \label{fig:fig1}
\end{figure}

The lattice translation symmetry in the direction of the effective magnetic field boundary $y=0$ preserves the quasi-momentum $q_x$, while motion in the $y$ axis direction is confined.
The corresponding spinor equation describing this confined motion reads
\begin{equation} \label{eq_cont_schroed}
    E \Psi = \hbar v_F [\tau_z \sigma_x (q_x-A_x) -i \sigma_y \partial_y ] \Psi.
\end{equation}
Eliminating either of the two sublattice components of the pseudospinor $\Psi=(\Psi_\text{A},\Psi_\text{B})$ yields uncoupled equations for each component:
\begin{equation}  \label{eq_ABcomponents}
    \left[ (E/\hbar v_F)^2 -(\tau_z q_x + B |y|)^2 + \partial_y^2 \pm B {sgn}{(y)} \right] \Psi_\text{A(B)} = 0,
    \end{equation}
which are equivalent to the 1D Schr\"odinger equation with the effective potential
\begin{equation}
    U_\textrm{A(B)}(y) = (\tau_z q_x + B |y|)^2 \mp B {sgn}(y).
\end{equation}

The eigenstates of Eq.~\eqref{eq_cont_schroed} are illustrated in Fig.~\ref{fig:fig1}.
The corresponding energy absolute values are shown in the left panel (a) as functions of the normalized magnetic field $B$ and lateral wave vector $q_x$.
The allowed direction of lateral motion is determined by the sign of the group velocity $v_x = \partial E / \partial q_x$ and thus by the three factors: (i) the valley index $\tau_z = \pm 1$, (ii) the sign of the energy $E$, and (iii) the sign of the magnetic field $B$.
In the practically relevant case, where the energy of the state is controlled with external resonant excitation, the optical signal propagation direction is bound to the valley, which can be employed for valley splitting of valley filtering.
All surfaces $|E/J|(Ba^2,\tau_zq_xa)$ coalesce in the limit $B\rightarrow 0$ at the red line with the slope corresponding to the Fermi velocity $v_F$.
Note that the strong coupling between the valley and the allowed transport direction is determined by the symmetry of the structure, quantified by the sign of $B$, and is thus preserved even in the limit $|B|\rightarrow 0$.

The corresponding field densities, separated into sublattice components, as well as the effective potentials of Eqs.~\eqref{eq_ABcomponents}, are shown in Fig.~\ref{fig:fig1}(b) for $Ba^2=-\tau_zq_xa=0.1$.
In the high-energy limit $|E|\gg\hbar v_FB$ the model is similar to a harmonic oscillator, therefore wave packets resembling coherent states are expected to exhibit oscillations in the $y$ direction.
These oscillations, combined with unidirectional lateral motion in the $x$ direction, result in snake-like wave packet center of mass trajectories.
In the low energy quantum limit, however, similar snake-like trajectories correspond to superpositions of the lowest energy states of Eq.~\eqref{eq_cont_schroed} of different parities, to which both even and odd index eigenstates contribute.

It should be noted that the property of unidirectional motion along $x$ axis (for fixed signs of valley index (i),  energy (ii), and magnetic field (iii)) is not limited to the states demonstrating oscillations along $y$ axis.
As an example, any eigenstate of Eq.~\eqref{eq_cont_schroed}, or a superposition of states with the same parity, exhibiting beating (visible as periodic signal broadening in real space) rather than oscillations, has a defined group velocity in the $x$ direction thus giving rise to the valley filter property.
Valley filtering feature is captured in the continuous model and is further illustrated in simulations of experimentally relevant setups.

We also note that in the classical limit $B\rightarrow0$, unidirectional motion is only allowed for $\tau_zq_x>0$, which is demonstrated by the cusp shape of the energy dispersion, highlighted with red in Fig.~\ref{fig:fig1}.
For positive values of $\tau_zq_x$, the dispersion slope corresponds to the Fermi velocity of the unstrained graphene structure.
In turn, the case $\tau_zq_x>0$ is characterised with a flat dispersion and corresponds to cyclotron motion of polaritons in effectively separated semiplanes $y<0$ and $y>0$.

It is illustrative to compare the predictions obtained within the continuous zero-energy approximation with that of the full time-dependent tight-binding model (TBM). We considered laterally infinite zigzag ribbon consisting of 72 zigzag chains of cavity pillars, see \textit{Supplementary materials}. The binding strength between pillars belonging to nearest chains at the structure center $y=0$ is equal to the coupling parameter $J$ within each chain. Magnitude of coupling decreases linearly with the distance from the center of the structure and vanishes at top and bottom edges. The described coupling gradient is equivalent to an effective field $B \sim 0.01a^{-2}$, for which the continuous Eq.~\eqref{eq_ABcomponents} predicts the energy gap $\Delta E \sim 0.1 J$, separating the ground state from the excited ones at the taken value of wave vector $q_x \sim 0.5 a^{-1}$ (measured from Dirac point $K$ as in the low-energy approximation).

\begin{figure}[!ht]
    \centering
    \hspace{9pt} $\Psi_0+\Psi_2$ \hspace{40pt} $\Psi_0+\Psi_1$ \hspace{40pt} Classical
    \includegraphics[width=8.5cm]{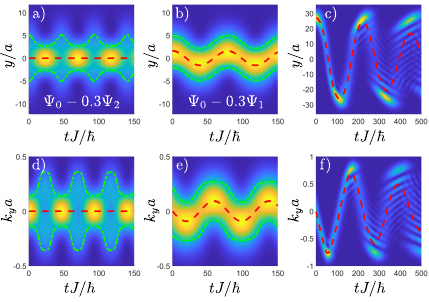}
    \caption{Tight-binding model for the snake state motion in the $y$ direction at fixed lateral wave vector $q_x=0.2 K$ in a structure composed by 72 zigzag chains. Panels (a-c) and (d-f) show the wave function evolution in real and reciprocal space respectively. The initial conditions are (a,d) superposition of the ground state and the second excited state, (b,e) superposition of the ground state and the first excited state, (c,f) Gaussian profile significantly shifted from the symmetry axis. Dashed red lines show the center of mass motion and green dot-dashed lines indicate the half-widths of wave packets.}
    \label{fig2}
\end{figure}

We simulated the dynamics of the field starting from three different sets of initial conditions, mimicking experimentally feasible Gaussian excitation wave-packets of selected width and shift from the structure symmetry axis $y=0$.
The results are shown in Fig.~\ref{fig2} both in real (panels a-c) and reciprocal space (panels d-f). The initial conditions were chosen to approximate superposition of the ground state $\Psi_0$ with the first (Fig.~\ref{fig2}a,d) and the second (Fig.~\ref{fig2}b,e) excited states $\Psi_1$ and $\Psi_2$ or as a Gaussian wave packet of width $w=4a$ centered at $y_0 = 27a$ (Fig.~\ref{fig2}c,f).

In accordance with expectations, the combination of the ground state and the second excited state (both are parity-symmetric) leads to oscillatory symmetric patterns both in real and reciprocal space. The combination of even ground state and odd first excited state wave functions results in snake-like behavior in the low-energy limit. Finally, excitation sufficiently shifted from the symmetry axis $y=0$ allows realizing classical snake states. The oscillation periodicity in the first and the second cases $T\sim50 \hbar/J$ is in agreement with the energy distance $\Delta E \sim 2\pi\hbar/T$ given by the continuous model.

\begin{figure}[ht!]
    \centering
    \includegraphics[width=\linewidth]{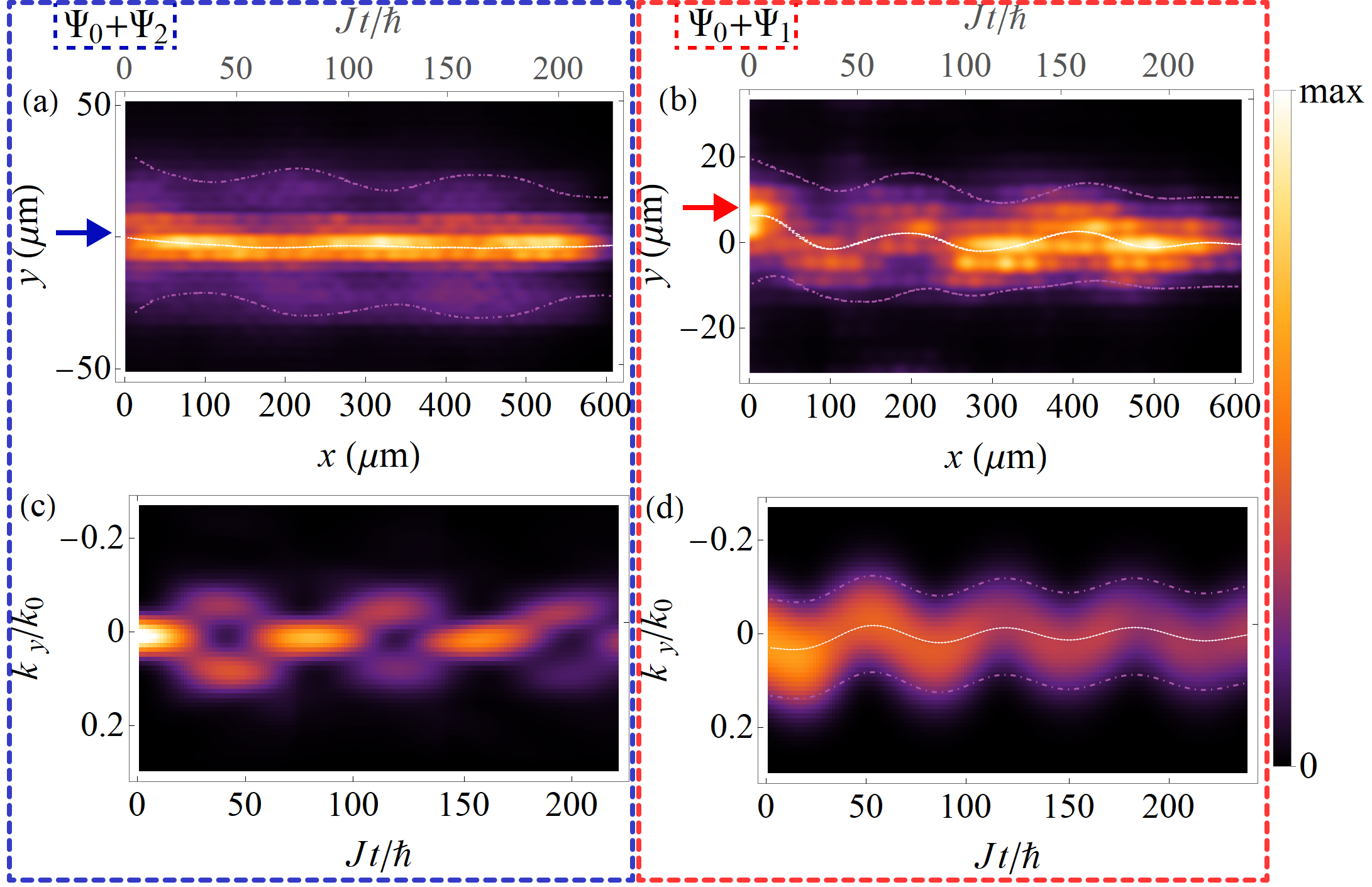}
    \caption{Wave packet propagation along the boundary between the domains of opposite-signed effective magnetic field. Initially the wave packet was excited symmetrically at the boundary for panels (a),(c), and with the center of Gaussian wave packet shifted from the boundary for panels (b) and (d), see arrows.
    Such initial states correspond to a combination of the ground state $\Psi_0$ and the second excited state $\Psi_2$ or the first one $\Psi_1$ respectively.
    The upper panels (a) and (b) are obtained from the integral intensity over the whole simulation time, which mimics time-integrated experimental intensity measurement.
    The $x$ coordinate axis is supplemented with the effective time axis using the mapping $t=x/c$, shown above panels (a),(b) to highlight the similarity with Fig.~\ref{fig2}a,c.
    The dashed lines show the half-maxima of the signal along vertical axis and the trajectory of the wave packet mass center.
    The lower panels show the time dependence of the wave packet profile along $y$ axis in the momentum space. The time is normalized on the characteristic timescale $\hbar/J$. Y-axis in the upper two panels demonstrates the real scale used in the modeling. The size in the $x$ direction is in correspondence with the one used in the tight-binding and continuous-wave calculations since the ribbon is periodic/infinite along the $x$-axis.}
    \label{fig3}
\end{figure}

\begin{figure}
    \centering
    \includegraphics[width=\linewidth]{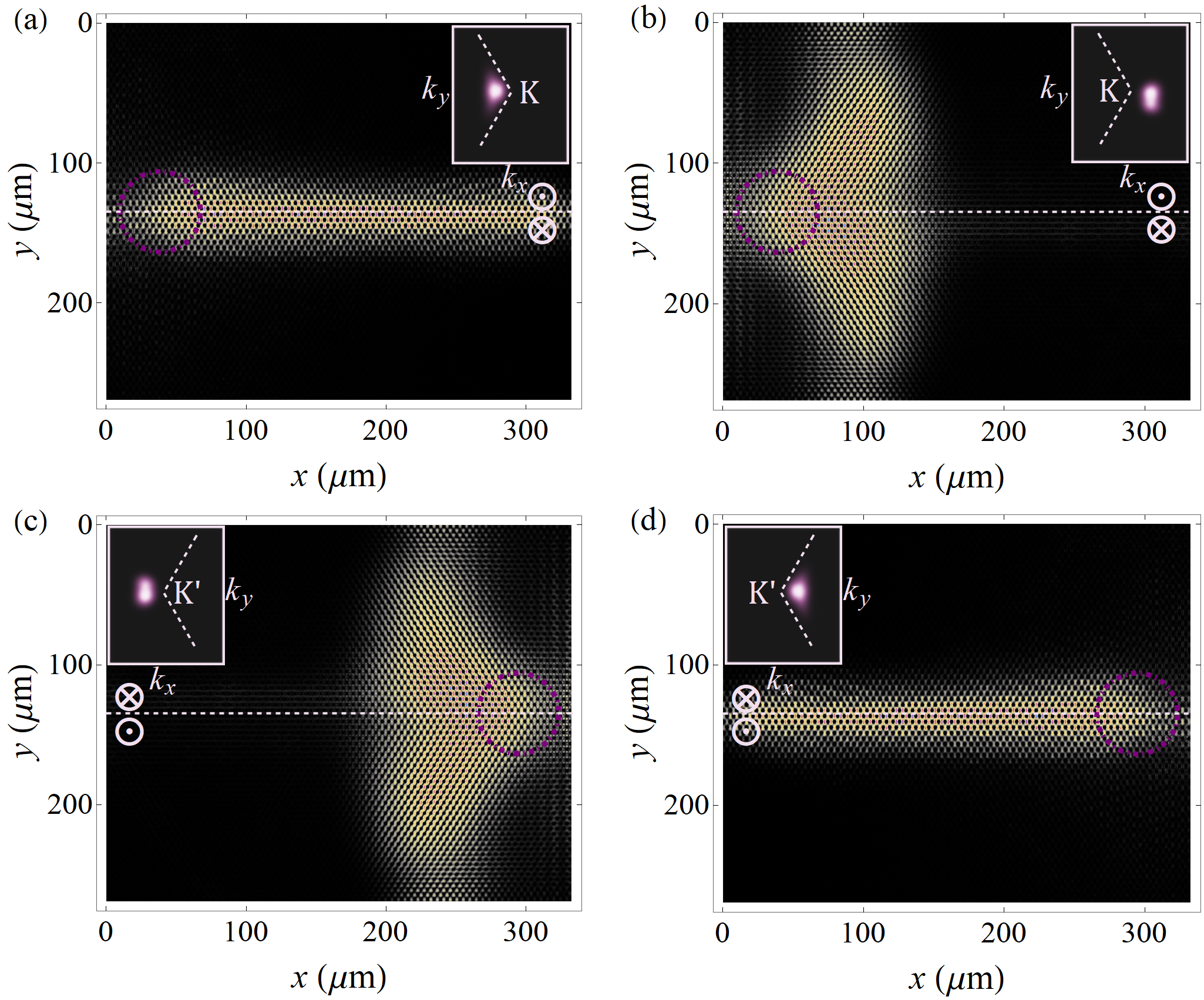}
    \caption{Demonstration of valley filtering with numerically solved Eq.~\ref{eq:SE}.
    Time integrated intensity of a wave packet excited at different points of the reciprocal space: (a) and (b) for the K point and (c), (d) for the $K^\prime$.
    The wave packet center is shifted from the Dirac points by $|q_x|=0.14K$, however, for panels (a) and (c) the quasi-wave vector $q$ is negative, while for panels (b) and (d) it is positive.
    Purple circles indicate the half-width of the initial Gaussian-type pulse. The insets show the corresponding integrated intensities in the reciprocal space.}
    \label{fig4}
\end{figure}

In addition, we performed detailed numerical simulations of the Schr{\"o}dinger equation for photonic graphene lattice, Eq.~(\ref{eq:SE}).
Potential $V(x,y)$ is symmetric with respect to horizontal mirror reflection axis $y=0$ and resembles the structure from Ref.~\cite{Jamadi2020} attached to its mirrored copy (see Fig.~\ref{fig:figs1}). Its shape provides coupling parameter gradients as in TBM resulting in the opposite directions of synthetic magnetic field in upper and lower half-spaces as required to observe the snake states (see Figs.~\ref{fig3},\ref{fig4}).

In order to quantitatively characterize the geometry of photonic honeycomb lattice with strain realized via coupling gradient along the $y$ direction, it is useful to introduce a parameter $\alpha = \delta d/d_0$ as a ratio of the change in the distance between the pillar centers in the nearest zigzag chains $\delta d$ per one vertical period to its value $d_0=2.7~\mu m$ in the undeformed part of the structure near the center. Pillar diameter 3.2~$\mu$m is chosen based on previously reported structures~\cite{Jamadi2020}.
We addressed two types of photonic graphene structures with different magnitudes coupling parameter gradient characterised by (i) $\alpha=0.014$ and (ii) $\alpha=0.056$.

The initial wave functions have a form of plane wave with wave vector $\mathbf{k} = (k_x,k_y)$, modulated by a Gaussian:
\begin{equation}\label{psi0}
    \Psi_{in} = \frac{1}{N} \exp\left[ -\frac{(x-x_0)^2}{2\sigma_x^2}-\frac{(y-y_0)^2}{2\sigma_y^2} + i(k_xx+k_yy) \right].
\end{equation}
In the expression above $N$ is the normalization constant, $\sigma_{x,y}$ characterize wave packet widths along the two main symmetry directions.
In the addressed configurations the initial wave vectors have only $k_x$ component with a value from the vicinity of the Dirac point $K=K_x=4\pi /(3 \sqrt{3} a)$, where $a$ is the lattice constant.

Figure~\ref{fig3} illustrates time evolution of a wave packet in photonic graphene ribbon for various initial conditions. The top row (Figs.~\ref{fig3}(a),(b)) shows the time-averaged wave function density in the real space, representing a trace left by a propagating wave packet and corresponding to time-integrated near-field polariton emission from the structure. The signal was filtered in space to mitigate the non-uniformity caused by discreteness of lattice and the original wave function with resolved pillars is presented in \textit{Supplementary materials}, Fig.~\ref{fig:figs2}.
As the group velocity of the wave packet along the $x$ direction is determined by the Fermi velocity of the Dirac cone, the $x$ coordinate may be mapped onto dimensionless time $Jt/\hbar \equiv Jx/(\hbar v_F)$ to facilitate comparison with the results of the tight-binding model shown in Fig.~\ref{fig2}.
The coupling parameter was estimated as $J=0.26$~meV.
The distinctive features of various type of behavior can be also obtained from momentum space measurements at different time moments. The bottom row (Figs.~\ref{fig3}(b),(d)) shows the wave function density in the reciprocal space, integrated over the $x$ dimension for each time step, as a function of normalized time, corresponding to the expected angular distribution of the emission in the far field.

In the first case (Fig.~\ref{fig3}(a),(c)), the wave packet is initially centered at the symmetry axis ($y_0=0$) and has the wave vector $q_x = - 0.2K$, which corresponds to $k_{x}\equiv K+q_x=0.8 K$.
While propagating in the $x$ direction, it exhibits oscillating width along the $y$ axis, which is illustrated with the calculated position of the density half-maximum (see the purple dash-dotted curve in Fig.~\ref{fig3}(a)).
The snake state motion is most vividly presented in Fig.~\ref{fig3}(b),(d).
In this case, the initial Gaussian wave packet is shifted from the symmetry axis by $ y_0 =11.5~\mu$m, its wave vector lying in the $x$ direction with $k_{x}=0.86K$.
The center of mass trajectory of the wave packet is shown with the white dashed line in Fig.~\ref{fig3}(b),(d) in both real and reciprocal spaces as a guide for an eye, as well as the positions of the density half-maxima along the $y$ axis.
As to have a well distinguishable beats only few eigenfunctions are to be excited, observed behavior is sensitive to the parameters of initial signal shape both in TBM and Schr\"odinger equation simulations.
Overall, the more detailed simulations reproduce the main features of the simplified tight-binding model shown in Fig.~\ref{fig2}(b),(e). 

Finally, we have simulated a scenario demonstrating valley filtering property of the proposed structure by numerical solving the Schr{\"o}dinger equation, Eq.~(\ref{eq:SE}), and within the tight-binding approximation (see \textit{Supplementary materials}).
Fig.~\ref{fig4} demonstrates the time integrated emission when the initial wave packet is excited at different valleys with two different wave vectors in each valley.
The initial Gaussian wave packets with the widths $w_{x,y}=50~\mu$m and in-plane carrier wave vector projections $k_x = \pm K \pm 0.14 K$ are centered at the symmetry axis $y_0=0$.
It is seen that only one propagation direction is allowed for the wave packet in each valley, which is consistent with the symmetry properties of the group velocity shown in Fig.~\ref{fig:fig1}.
%


\textit{In conclusion}, we are proposing a new type of optical confined states with distinct symmetry properties, emerging in symmetrically strained photonic graphene structures. The observed states naturally appear at the boundary between the domains with opposite directions of synthetic magnetic field thus sharing similarity with electron snake states.

We addressed such photonic graphene structures with models of increasing complexity and precision: continuous approximation in the low-energy limit and discrete realization of tight-binding model as well as by solving the two-dimensional Schr\"odinger equation for exciton-polaritons in the potential corresponding to properly etched microcavity.
Proposed optical snake states can be experimentally revealed be measuring polariton emission using pulsed excitation of wave packets with specific fingerprints both in real and momentum space.

We note that, in contrast to topological edge states protected
by time-reversal symmetry breaking and nonzero Chern numbers, optical snake states are only unidirectional for wavepackets
excited in a given valley, rendering it phenomenologically similar to quantum valley Hall effect. In addition, no energy gap
opens, which can be seen from Fig.~\ref{fig:fig1}, summarizing predictions
of the continuous model, accurate in the vicinity of Dirac points.

The proposed photonic graphene symmetric structure with opposite strain directions can be used for valley filtering as the allowed direction of propagation along the symmetry axis is determined by the valley and the energy sign.
Although the structure spectrum as a whole is topologically trivial according to the correspondence between the topology and symmetry, each valley is locally time-reversal asymmetric, which results in unidirectional valley transport due to valley Hall effect.
In contrast to spectrally narrow bands of topologically protected edge states, snake states responsible for valley filtering exist in a broad frequency range, limited by the structure size rather than the bulk stop-band width.

\textit{Acknowledgement}. 
O.B. and S.K. acknowledge the financial support from the Institute for Basic Science (IBS) in the Republic of Korea through the project IBS-R024-Y3. This work was partly supported through the project IBS-R024-D1 (O.B.). A.N. acknowledges support by the Russian Science Foundation under Grant No. 22-12-00144 and MIPT Priority-2030 Program. E. C. acknowledges the Basis foundation, grant No. 21-1-3-30-1. O.B. thanks D.~Solnyshkov and G.~Malpuech for helpful discussions.

\bibliography{references}

\clearpage

\setcounter{equation}{0}
\setcounter{figure}{0}
\setcounter{table}{0}
\setcounter{page}{1}
\makeatletter
\renewcommand{\theequation}{S\arabic{equation}}
\renewcommand{\thefigure}{S\arabic{figure}}
\renewcommand{\thetable}{S\arabic{table}}
\renewcommand{\thesection}{S\Roman{section}}
\renewcommand{\thepage}{S\arabic{page}}
\renewcommand{\bibnumfmt}[1]{[S#1]} 
\renewcommand{\citenumfont}[1]{S#1}

\onecolumngrid

\begin{center}
\textbf{\large Supplemental Material: Optical snake states in photonic graphene}
\end{center}

\section{Tight-binding model of photonic graphene} \label{S1}

We assume the planar cavity Hamiltonian with the lattice potential
\begin{equation}
    H = - \frac{\hbar}{ 2m} (\partial_x^2 + \partial_y^2) + \sum_l V_0(\bm{r}-\bm{r}_l),
\end{equation}
where $r_l$ are the lattice vertex positions and $V_0(\bm{r})$ is the trapping potential emulating photon confinement in a cylindrical vertex.
For the ground state, producing the lowest energy band of the lattice, one has
\begin{equation} \label{eq:Psi0}
    \left[-\frac{\hbar^2 }{ 2m}(\partial_x^2+\partial_y^2) + V_0(\bm{r}) \right] \Psi_0(\bm{r}) = E_0 \Psi_0(\bm{r}).
\end{equation}
Substituting $\Psi(\bm{r}) = \sum_j \psi_j\Psi_0(\bm{r} - \bm{r}_j)$ in Eq.~(2), using \eqref{eq:Psi0}, multiplying by $\Psi_0(\bm{r}-\bm{r}_k)$ and integrating over the plane we get:
\begin{equation}
    i \hbar \frac{d \psi_k}{ dt} = \sum_{j\neq k} \psi_j \int \Psi_0^*(\bm{r}-\bm{r}_k) \sum_{l\neq j}V(\bm{r}-\bm{r}_l)\Psi_0(\bm{r}-\bm{r}_j) d \bm{r}.
\end{equation}
In the nearest neighbor approximation, the above expression reduces to
\begin{equation}
    i \hbar \frac{d \psi_k}{ dt} = -J\sum_{<jk>}\psi_j,
\end{equation}
where $J = -\sum_{l\neq j}\int \Psi_0^*(\bm{r}-\bm{r}_k) V_0(\bm{r}-\bm{r}_l) \Psi_0(\bm{r}-\bm{r}_j)$ and $<kj>$ denotes nearest neighboring vertex indices.

Introducing the two triangular sublattices A and B and the corresponding plane wave anzats $\psi_{A/B,j \bm{k}} = \psi_{A/B} \exp(i\bm{k}\bm{r}_{j,A/B})$, we arrive to the standard graphene Bloch Hamiltonian in the basis of the two sublattices:
\begin{equation}
    H_k = -J \left[ \begin{matrix}
        0 & f_{\bm{k}} \\
        f_{\bm{k}}^* & 0
    \end{matrix} \right],
\end{equation}
where $f_k = \sum_{j=1}^3 \exp(i\bm{k}\bm{r_j})$ with $r_j$ being the 3 displacement vectors, describing hopping from any A sublatice vertex to neighboring B sublattices vertices.
In the vicinity of the Dirac points $K$ and $K'$, expanding $f_{\bm{k}}$ to the leading order results in the low-energy approximation Hamiltonian, Eq.~(3).

\section{1D tight-binding model calculations}

\begin{figure}[ht!]
    \centering
    \includegraphics[width=0.5\linewidth]{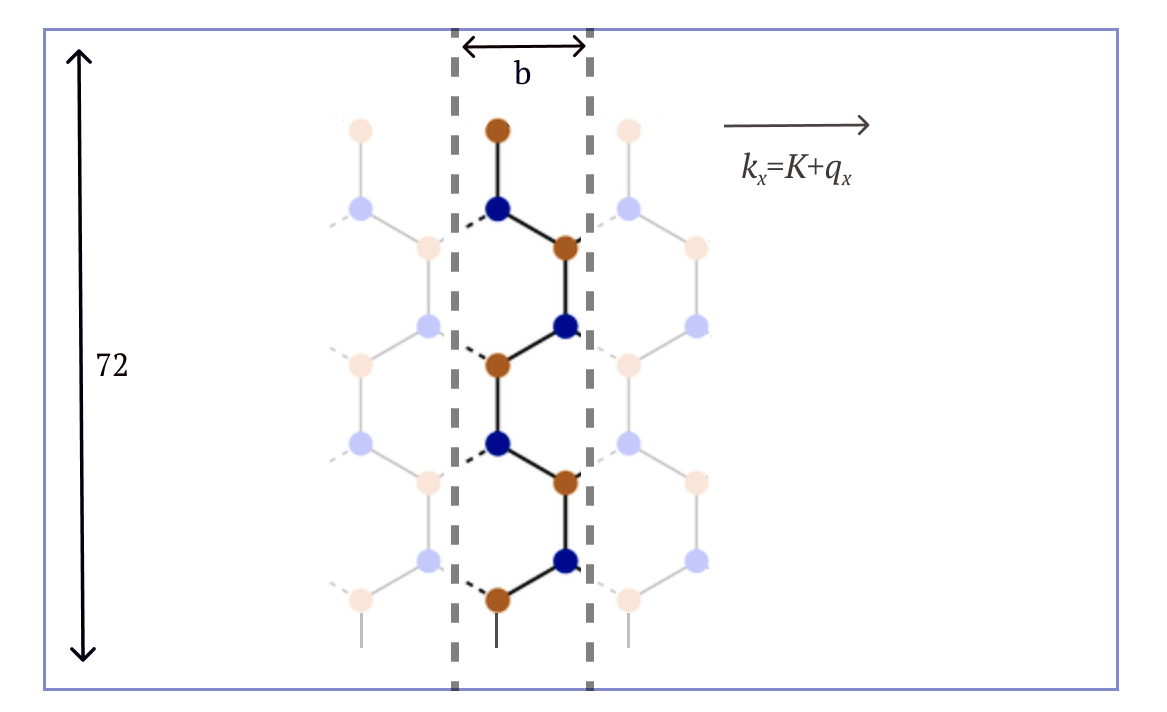}
    \caption{For TBM calculations, we use graphene ribbon formed of 72 horizontal infinite zigzag chains. The structure is periodic in $x$ direction with the unit cell formed of one vertical armchair chain denoted with dashed lines.}
    \label{fig:figs5}
\end{figure}

The structure of graphene ribbon for used TBM calculations is shown in Fig.~\ref{fig:figs5}.

The equation used in the time-dependent TBM calculations reads
\begin{equation}
    i \hbar \partial_t \psi_i = \sum_{\{j\}} t_{ij} \psi_j \exp(i(K+q_x)b\phi_{ij}),
\end{equation}
where $\psi_i$ is the wave function at site $i$, $t_{ij}$ is the hopping parameter (depending on $y$ coordinate for vertical bonds), $j$-summation is conducted over the 3 nearest neighbors of the site $i$ and phase factor $\phi_{ij}=-1,0,+1$ depends on the indices of 1D unit cells containing $i$-th and $j$-th sites. Due to linearity, the eigenproblem for the Hamiltonian from Eq. (S1) was firstly solved and then initial conditions were projected on the eigenstates to trace further time evolution.

\section{Parametrization of photonic lattice with effective strain}

In this section we provide more details on the coordinate potential profiles used in numerical solving of the Schr\"odinger equation, Eq.~(\ref{eq:SE}) of the main text. 

\begin{figure}[ht!]
    \centering
    \includegraphics[width=0.55\linewidth]{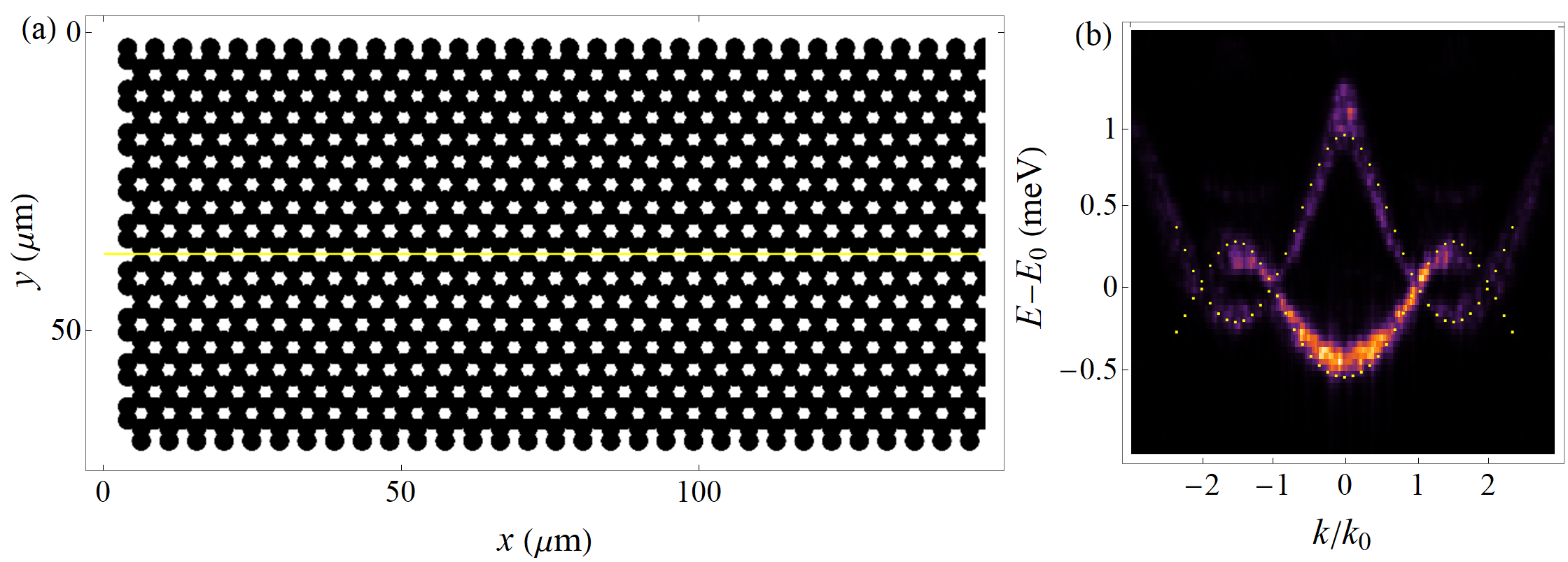}
    \caption{(a) Strained photonic graphene lattice (cutted in the $x$ direction for better representation, $L_y=70~\mu$m) used for numerical simulations of the Schr\"odinger equation (see Eq.~(1) of the main text). For this lattice the effective magnetic field strength is described by $\alpha=0.056$ (as for the one of snake states, Fig.~4). The yellow line marks zero deformation one (border between two parts with the opposite tension.) (b) numerically calculated dispersion for the considered strained photonic graphene ($s$-bands). $E_0=1.56$eV. Yellow dots correspond to the one obtained within the tight-binding model.}
    \label{fig:figs1}
\end{figure}

Figure~\ref{fig:figs1} demonstrates a photonic honeycomb lattice with strain obtained via variation of hopping parameter controlled. For this lattice, the strain is arranged in a way that the vertical spacing between the neighbouring, i.e. belonging to the nearest zigzag chains, pillar centers $d$ decreases when approaching boundaries compared to the center marked by the thick yellow line. In the center $d=d_0$, where $d_0$ labels the distance between pillars for the underformed structure. Such a lattice, which is characterized by interpillar distance gradient $\alpha=0.056$, is used to obtain the oscillations of the wave packet center of mass resulted in the snake states (see the main text and Fig.~\ref{fig3}(b),(d), and also Fig.~\ref{fig:figs2}). For the initial condition, the wave packet with the wave vector only in the $x$ direction, $k_x=0.86K$, and the center shifted from the symmetry line by $y_0=11.5~\mu$m, is characterized by non-equal widths in $x$ and $y$ directions: $w_x=20.25~\mu$m, $w_y=12.25~\mu$m. Here by the width $w$ of a Gaussian we mean the full width at half maximum (FWHM) which is related to the variance as $w=2\sqrt{2\text{ln}2}\sigma$, see Eq.~(\ref{psi0}) of the main text. To suppress unnecessary high-wave vector harmonics in the simulations, the initial wave function was additionally multiplied by Bessel function $J_0(|\mathbf{r}|,c_1)$ profiles centered at each pillar representing the $s$-state at isolated pillar (parameter $c_1$ provides proper zero-boundary conditions).

Further, a slightly more complex and bigger in size lattice is utilized in order to obtain the results showed in Fig.~\ref{fig3}(a),(c) of the main text (see also Fig.~\ref{fig:figs2}). It is created in a manner that $d<d_0$ on the center line, and described by $\alpha=0.014$. 

Additionally, Figure~\ref{fig:figs2} shows full (without additional filtering and compression in lateral direction) time-averaged wave function densities from which Figs.~\ref{fig3}(a),(b) have been obtained. Also, it worth to mention that since the emergence of the oscillatory behavior (for snake states or beats) is related to dominance of the ground state wave function $\Psi_0$ and first exited ones ($\Psi_1$ or $\Psi_2$) for an initial superposition state, it becomes sensitive to some system parameters as the effective magnetic field strength $\alpha$, wave vector $k_x$, shift of the initial Gaussian wavefunction center from the symmetry axis $y_0$, as well as its width $w$. 

For integration the Schr\"odinger equation, the third order Adams–Bashforth scheme with time step $\Delta t=0.8\times 10^{-4} \hbar/J$ $(2\times 10^{-4} ps)$ and $2048\times 2048$ mesh grid with spatial step $0.2~\mu m$ were used. The polariton mass was taken $m=5\times 10^{-5}m_e$ and 3~meV was the height of the potential barrier.

\section{Valley filtering property}

\begin{figure}[!ht]
    \centering
    \includegraphics[width=8.5cm]{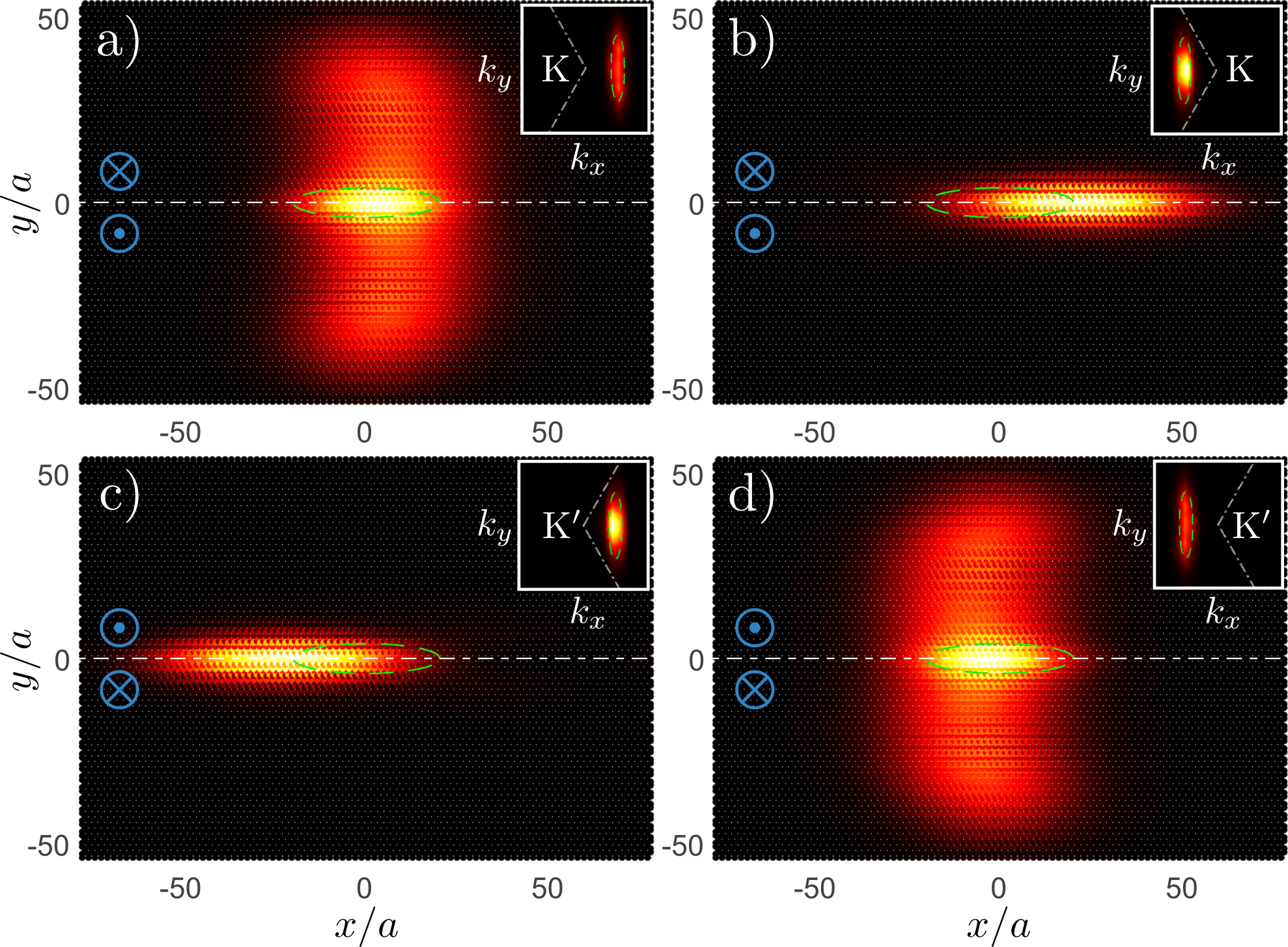}
    \caption{Full 2D tight-binding simulation of valley filtering using optical snake states.
    Panels show density of wave packets, time-integrated from $t=0$ to $t=30 \hbar/J$.
    The initial states of the wave packet simulate excitation with oblique Gaussian pulses (half-width is shown with green dashed lines) at the K (panels a and b) and K$^\prime$ (panels c and d) valleys.
    For each valley, only one direction of wave guiding in the $x$ direction is allowed.
    Insets show corresponding time-integrated density in the reciprocal space.
    The directions of effective magnetic field are shown with blue.
    }
    \label{fig:figs3}
\end{figure}

\begin{figure}[ht!]
    \centering
    \includegraphics[width=\linewidth]{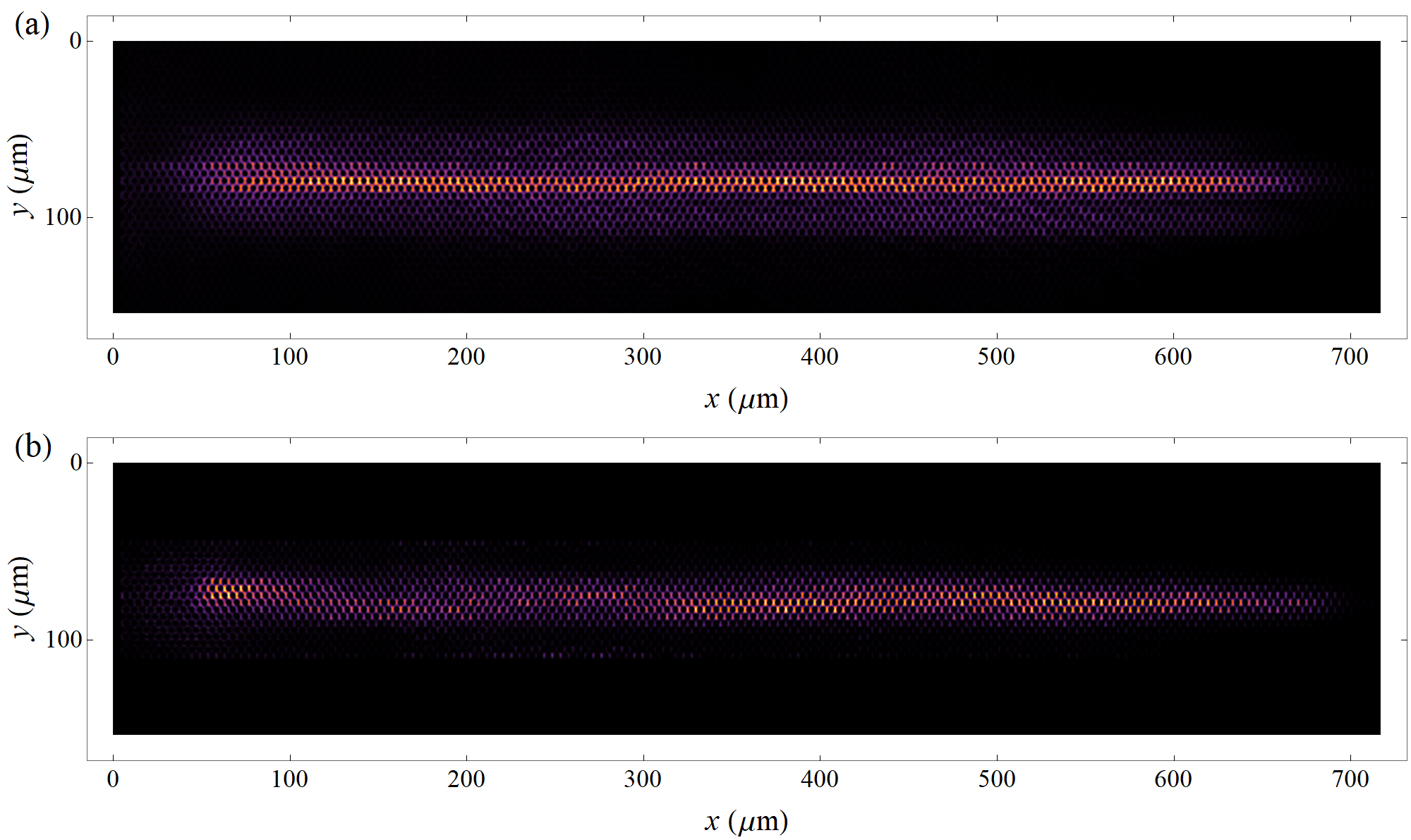}
    \caption{Integrated intensity over time for the cases of quantum beats (a) and snake states (b). }
    \label{fig:figs2}
\end{figure}

In order to solidify discussion on the valley filtering properties of strained photonic lattices, here we present results obtained in simulation based on the 2D TBM approach. 
Figure~\ref{fig:figs3} shows density snapshots of wave packets (of a Gaussian shape $t=0$) with horizontal and vertical half-widths $w_x=20a$, $w_y=4a$, and with in-plane carrier wave vector projections $k_x = \pm K \pm 0.1 K$, centered at the symmetry axis $y=0$.
%

%
\begin{figure}[ht!]
    \centering
    \includegraphics[width=\linewidth]{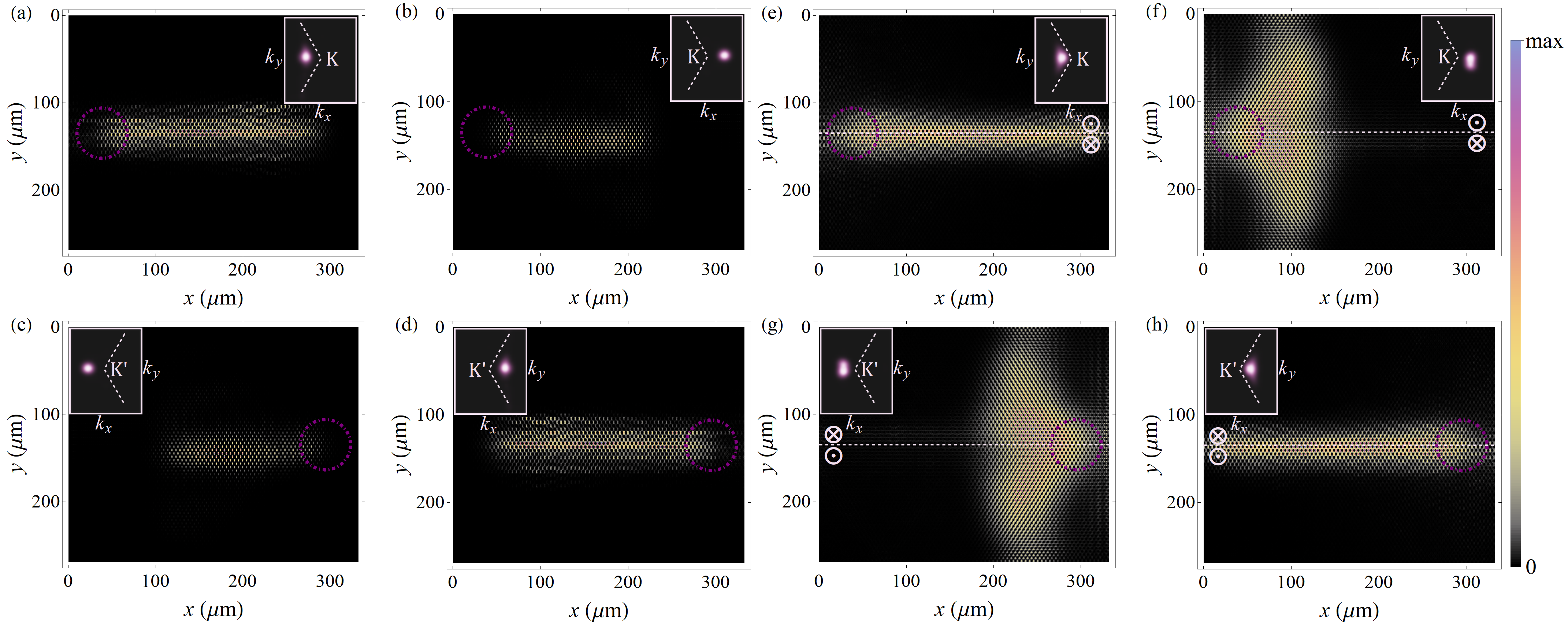}
    \caption{Demonstration of the valley filtering feature of the strained photonic graphene. (a)-(d) for underformed lattice. (e)-(h) the same as Fig.~\ref{fig4} of the main text. }
    \label{fig:figs4}
\end{figure}

To emphasize valley filtering properties of strained photonic graphene, we compare a wave packet propagation for the lattice with and without existence of the effective magnetic field. Figure~\ref{fig:figs4} shows time-averaged wave function densities obtained by numerical solving of the Schr\"odinger equation, Eq.~\ref{eq:SE} of the main text, for undeformed lattice (a)-(d) and in the presence of the strain (e)-(h), see also Fig.~\ref{fig4} of the main text.
The width of the initial wavepacket is $w=50~\mu$m (marked by purple dashed circles), $k_x=\pm K\pm 0.14K$. The insets show time-integrated intensities in the momentum space. The white dashed line in Fig.~\ref{fig:figs4}(e)-(h) corresponds to the lattice symmetry axis $y=0$. The directions of effective magnetic field are also shown.

%

\end{document}